\documentclass[a4paper,fleqn]{cas-dc}

\usepackage[authoryear]{natbib}
\usepackage{textcomp}
\usepackage{amsmath,amssymb,amsfonts}
\usepackage{bm}
\usepackage{makecell}
\usepackage[section]{placeins}

\usepackage{cleveref}
\crefname{figure}{fig.}{figures}
\Crefname{figure}{Fig.}{Figures}

\usepackage{pifont}
\newcommand{\Checkmark}{\ding{51}}     
\newcommand{\XSolidBrush}{\ding{55}}   

\usepackage{placeins}   
\usepackage{float}      

\makeatletter
\def\fps@figure{htbp!}
\def\fps@table {htbp!}

\makeatother

\begin{document}
\newcommand{\ourmodel}{MTRL-FIE}
\newcommand{\encoder}{MFE}
\newcommand{\decoder}{SHD}
\newcommand{\attention}{TFA}
\let\WriteBookmarks\relax
\def\floatpagepagefraction{1}
\def\textpagefraction{.001}
\let\printorcid\relax

\flushbottom

\shorttitle{Multi-Scale Target-Aware Representation Learning}
\shortauthors{Haofan Wu et~al.}

\title [mode = title]{Multi-Scale Target-Aware Representation Learning for Fundus Image Enhancement}

\author[1,2,3]{Haofan Wu}
\credit{Writing – original draft, Conceptualization, Data curation, Methodology, Investigation, Visualization}
\fnref{equal}
\author[1,3]{Yin Huang}
\credit{Writing – review and editing, Conceptualization, Methodology, Supervision, Project administration}
\fnref{equal}
\author[1,2]{Yuqing Wu}
\credit{Data curation, Formal analysis, Software}
\fnref{equal}
\author[1,2]{Qiuyu Yang}
\credit{Formal analysis, Visualization}

\author[1,2]{Bingfang Wang}
\credit{Writing – review and editing, Validation}

\author[4]{Li Zhang}
\credit{Supervision}

\author[5]{Muhammad Fahadullah Khan}
\credit{Resources}
\author[5,6]{Ali Zia}
\credit{Resources}
\author[5]{M. Saleh Memon}
\credit{Resources}
\author[5]{Syed Sohail Bukhari}
\credit{Resources}
\author[5]{Abdul Fattah Memon}
\credit{Resources}
\author[1,2]{Daizong Ji}
\credit{Funding acquisition, Resources}
\author[1,2]{Ya Zhang}
\credit{Supervision, Visualization, Project administration, Funding acquisition, Resources}
\author[7]{Ghulam Mustafa}
\credit{Resources, Methodology, Software}
\author[1,2,3]{Yin Fang}
\cormark[1]
\ead[E-mail address]{yin_fang@tongji.edu.cn}
\credit{Writing – review \& editing, Funding acquisition, Project administration, Resources, Supervision}
\affiliation[1]{organization={Research Center for Translational Medicine, Medical Innovation Center and State Key Laboratory of Cardiology, Shanghai East Hospital},            
                city={Shanghai},
                postcode={200120},                 
                country={China}}
\affiliation[2]{organization={The Institute for Biomedical Engineering \& Nano Science, Tongji University School of Medicine},            
                city={Shanghai},
                postcode={200120},                 
                country={China}}
                
\affiliation[3]{organization={Shanghai Research Institute for Intelligent Autonomous Systems},            
                city={Shanghai},
                postcode={201210},                 
                country={China}}      

\affiliation[4]{organization={Department of Ophthalmology, Tongji Hospital, School of Medicine, Tongji University},     
                city={Shanghai},
                postcode={200000},                 
                country={China}}      

\affiliation[5]{organization={Ophthalmology, Isra Postgraduate Institute of Ophthalmology},            
                city={Karachi},
                postcode={},                 
                country={Pakistan
}}      

\affiliation[6]{organization={The Eye Centre, 
South City Hospital},    
                city={Karachi},
                postcode={},                 
                country={Pakistan
}}      
\affiliation[7]{organization={Center for the Development of Laboratory Equipment, Pakistan Council of Scientific and Industrial Research},            
                city={Karachi},
                postcode={},                 
                country={Pakistan
}}

\cortext[cor1]{Corresponding author.}
\fntext[equal]{Equal contribution.}

\begin{abstract}
High-quality fundus images provide essential anatomical information for clinical screening and ophthalmic disease diagnosis.
Yet, due to hardware limitations, operational variability, and patient compliance, fundus images often suffer from low resolution and signal-to-noise ratio. 
Recent years have witnessed promising progress in fundus image enhancement.
However, existing works usually focus on restoring structural details or global characteristics of fundus images, lacking a unified image enhancement framework to recover comprehensive multi-scale information.
Moreover, few methods pinpoint the target of image enhancement, e.g., lesions, which is crucial for medical image-based diagnosis.
To address these challenges, we propose a multi-scale target-aware representation learning framework (\ourmodel{}) for efficient fundus image enhancement.
Specifically, we propose a multi-scale feature encoder (\encoder{}) that employs wavelet decomposition to embed both low-frequency structural information and high-frequency details.
Next, we design a structure-preserving hierarchical decoder (\decoder{}) to fuse multi-scale feature embeddings for real fundus image restoration.
\decoder{} integrates hierarchical fusion and group attention mechanisms to achieve adaptive feature fusion while retaining local structural smoothness.
Meanwhile, a target-aware feature aggregation (\attention{}) module is used to enhance pathological regions and reduce artifacts.
Experimental results on multiple fundus image datasets demonstrate the effectiveness and generalizability of \ourmodel{} for fundus image enhancement.
Compared to state-of-the-art methods, \ourmodel{} achieves superior enhancement performance with a more lightweight architecture.
Furthermore, our approach generalizes to other ophthalmic image processing tasks without supervised fine-tuning, highlighting its potential for clinical applications. The code and models will be made publicly available at https://github.com/RyanWu31/MTRL-FIE.
\end{abstract}

\maketitle

\section{Introduction}
Fundus imaging provides rich anatomical information and is widely used for screening and diagnosing ophthalmic diseases such as glaucoma, diabetic retinopathy, and cataracts \citep{li2021applications}. 
However, uncertainties in acquisition equipment, physician practices, and patient cooperation introduce large variations in fundus image \citep{philip2005impact}.
These variations degrade image quality, causing defocus or motion blur, artifacts, and improper exposure.
Degraded fundus images not only jeopardize clinical diagnosis and screening, but also significantly impede the development of image-based computer-aided diagnosis \citep{deng2022rformer}.
For example, Google's team found that their diabetic retinopathy AI diagnostic system failed to recognize over one-fifth of the images due to insufficient image clarity \citep{google_ai_diabetic_retinopathy}.
Various image enhancement algorithms, including histogram equalization  \citep{setiawan2013color}, spatial filtering  \citep{cheng2018structure}, and frequency filtering  \citep{cao2020retinal,mitra2018enhancement}, have been applied to improve fundus image quality and assist in ophthalmic disease diagnosis.
However, these traditional imaging enhancement approaches are sensitive to noise and rely on prior knowledge, often leading to loss of structural details.

In recent years, the development of deep learning has greatly promoted research into image enhancement. 
These methods can be categorized into two types: synthetic data-based methods and domain adaptation methods.
Synthetic data-based enhancement methods leverage paired datasets of low-quality and high-quality images to learn enhancement priors  \citep{deng2022rformer,li2021applications,cheng2018structure,shen2020modeling,csevik2014identification,perez2020conditional,li2022annotation,liu2022degradation}. These methods are easy to train, and the required training data is relatively easy to obtain.
However, they struggle to address the domain gap between synthetic and real-world low-quality fundus images, leading to poor generalizability in clinical settings.
Domain adaptation methods  \citep{guo2023bridging,li2022annotation,ma2021structure} attempt to improve generalization by accessing test data. This strategy can enhance the transferability from synthetic to real-world images. However, it often involves the joint training of multiple networks, making the training process more complex and unstable.

Despite the progress made by these approaches in restoring structural details and reducing noise impacts, they overlook two critical issues:
\textit{One is that most deep learning-based approaches often enhance images at only a single scale, lacking a unified multi-scale learning.}
The single-scale image enhancement leads to the loss of local details  \citep{zhang2018unreasonable,johnson2016perceptual} and overemphasis of global features  \citep{vedaldi2012efficient}.
For example, Convolutional Neural Network (CNN) based methods, limited by fixed receptive fields, struggle to enhance global and local characteristics simultaneously, resulting in an unclear enhancement objective  \citep{huang2017densely}. 
Transformer-based methods  \citep{deng2022rformer} effectively capture global information but exhibit limited sensitivity to local high-frequency features. 
Furthermore, frequency-based method  \citep{li2023generic} attempts to enhance images by directly utilizing high-frequency components, but lacks a comprehensive global modeling strategy. They fail to consistently enhance lesion boundaries, affecting clinical interpretability  \citep{guo2016lime}.  
Since pathological lesions and anatomical structures in fundus images are distributed across multiple spatial scales—from fine micro-vessels to large optic disc and macular regions—multi-scale learning is particularly beneficial for enhancing both global visibility and local diagnostic details.
\textit{Another issue is that these methods rely on the model to autonomously learn how to enhance images, which may cause non-uniform enhancement and undesired artifacts. }
For instance, GAN-based methods \citep{cheng2021secret} may introduce unrealistic pathological features, increasing the risk of misdiagnosis  \citep{ma2024review}.
Retinex-based methods \citep{dosovitskiy2020image,ronneberger2015u} often yield inconsistent enhancement results due to individual anatomical variations. \citep{jobson1997multiscale}.

To address these issues, we propose a multi-scale target-aware representation learning framework (\ourmodel{}) to incorporate comprehensive multi-scale features for efficient fundus image enhancement.
\ourmodel{} consist of a multi-scale feature encoder (\encoder{}) and a structure-preserving hierarchical decoder (\decoder{}).
Specifically, \encoder{} first employs wavelet transform to perform multi-scale decomposition of the input image, mining both low-frequency structural information and high-frequency details to achieve a unified enhancement.
Meanwhile, to further reinforce local details, the high-frequency features are further refined through depthwise separable convolution, which facilitates the decoder to restore key target. 
Next, \decoder{} is used to dynamically enhance structural information and fine detail by combining group attention and hierarchical feature fusion mechanisms.
Finally, we design a \attention{} module to ensure the target-aware features of high-frequency pathological regions, frequency-domain pathological regions, and structural characteristics in the original domain, effectively addressing unclear enhancement objectives and reducing image artifacts.
Notably, \ourmodel{} possesses a highly lightweight parameter size that supports its greater clinical applicability.

Our contributions in this work are several folds:
\begin{enumerate}
\item We propose \ourmodel{}, a lightweight self-supervised learning framework that effectively enhances structural information and fine-grained characteristics of fundus images.
\item We design a multi-scale feature encoder, which explicitly embeds both low-frequency structural information and high-frequency detail.
\item We develop a structure-preserving decoder to dynamically balance multi-scale features for retaining local structural smoothness, while introducing a \attention{} module to achieve target-aware features of crucial regions.
\item Extensive experiments demonstrate the superior performance and efficiency of our model.
\end{enumerate}

\section{Related Work}
\subsection{Fundus Image Enhancement}

Histogram equalization  \citep{mitra2018enhancement,joshi2008colour}, low-pass spatial filtering \citep{cao2020retinal}, and guided image filtering  \citep{cheng2018structure} are the earliest techniques for fundus image enhancement.
While these methods treat the entire image uniformly, which limits their ability to preserve details.
Recently, many deep learning-based image enhancement methods have emerged, which comprise more complex network structures capable of modeling rich characteristics.
For example, 
CycleGAN \citep{zhu2017unpaired} and StillGAN  \citep{ma2021structure} map non-uniform illumination domains to a high-quality domain, which restores fundus images from real, unpaired low-quality, and high-quality images.
In addition, they incorporate high-frequency feature supervision to ensure the consistency of fine-grained structural details \citep{yang2023retinal}.
I-SECRET \citep{cheng2021secret} employs contrastive learning to enhance fundus images. 
GVS \citep{zhang2022harmonizing} replaces the classifier with a segmentor, enabling more targeted lesion restoration.
These methods are modeled using unpaired data, which may generate unrealistic lesion features.
Another approach learns from paired low- and high-quality data for image enhancement.
For example, \cite{luo2020dehaze} train their model using synthetic cataract images.
Cofe-Net \citep{shen2020modeling} leverages vessel segmentation outputs in fundus image enhancement to preserve retinal structures.
HQG-Net \citep{he2023hqg} employs a variational information normalization strategy to make the enhanced images more conducive to automatic diagnosis.
Nevertheless, these methods depend on the model to independently learn enhancement targets, which may limit their practicability in clinical scenarios. 
As a result, domain adaptation has been introduced to solve this problem, such as maximum mean discrepancy  \citep{long2017deep} and domain discriminators  \citep{ganin2015unsupervised,tzeng2017adversarial}.
MAGE-Net  \citep{guo2023bridging} employs a mean-teacher training strategy to bridge the gap between synthetic and real fundus image enhancement.
SAME \citep{li2023enhancing} also employs a teacher–student knowledge distillation framework and validates its effectiveness across a broader range of imaging domains.
DASQE \citep{hou2024collaborative} employs a patch-level domain partitioning strategy that decomposes an image into multiple semantic components, allowing the framework to work entirely without reliance on high-quality images.
MDA-Net \cite{guo2024multi} leverages dynamic filters and degradation representations to enable the model to adapt to multi-domain data.
However, these domain adaptation approaches typically require the joint training of multiple networks, increasing both the complexity and instability of the training process.

\subsection{Encoder-Decoder Network-based Medical Image Enhancement}
Encoder-decoder Network have gained significant attention in medical image enhancement. 
U-Net was originally designed for medical image segmentation. This architecture has been adapted for image enhancement tasks  \citep{chen2022aau,isensee2021nnu,huang2020unet,ronneberger2015u,ibtehaz2023acc,lou2021dc}.
Variants of U-Net, such as U-Net++  \citep{zhou2018unet++}, Attention U-Net  \citep{oktay2018attention}, and Dense U-Net  \citep{li2018h}, further improve the feature extraction and enhancement capabilities by integrating attention mechanisms and dense connections.
Autoencoder-based methods have also been explored for medical image enhancement. These methods learn a compressed representation of the input image and reconstruct an enhanced version using a decoder network. Variational autoencoders and adversarial autoencoders  \citep{makhzani2015adversarial,rosca2017variational,zhao2018adversarially} have been utilized to enhance medical images while maintaining their anatomical structures.
Recently, transformer-based networks have been introduced to medical image processing, offering improved global context modeling. 
Vision Transformers (ViTs) and Swin Transformers have demonstrated their potential in medical image enhancement by leveraging self-attention mechanisms to capture global relationships and image details.

Hybrid models combining CNNs and transformers \citep{liang2021swinir,xu2024swin,jiang2024mutual,shao2022llu,zamir2022restormer,zhang2023novel} have further improved the enhancement performance by balancing local and global feature learning. These methods achieve remarkable image enhancement performance by integrating local and global information, but usually require substantial computational resources. MSQNet \citep{hou2025pathology} preserves original pathological features by combining multi-color space representations with a pathology-preserving Transformer in a semi-supervised framework.

Wavelet transform (WT) \citep{birkhoff2013cbms} is a multi-scale time-frequency analysis method that captures local features at different signal scales. It has been widely used in image compression, edge detection, and denoising \citep{mallat1989theory}. In deep learning, WT has been applied to enhance feature representations \citep{liu2018multi}, improve the quality of generated images \citep{guth2022wavelet}, and integrate with CNNs in various ways—for instance, by replacing certain convolution operations or compressing feature maps \citep{finder2024wavelet}. 
WT provides an invertible, parameter-sharing downsampling mechanism that reduces computational cost while preserving high-frequency structural details. 
This property makes it particularly suitable for medical imaging tasks where detail preservation and edge sensitivity are crucial.

\section{Method}
\subsection{Architecture Overview}
\begin{figure}[!t]
\centerline{\includegraphics[width=\columnwidth]{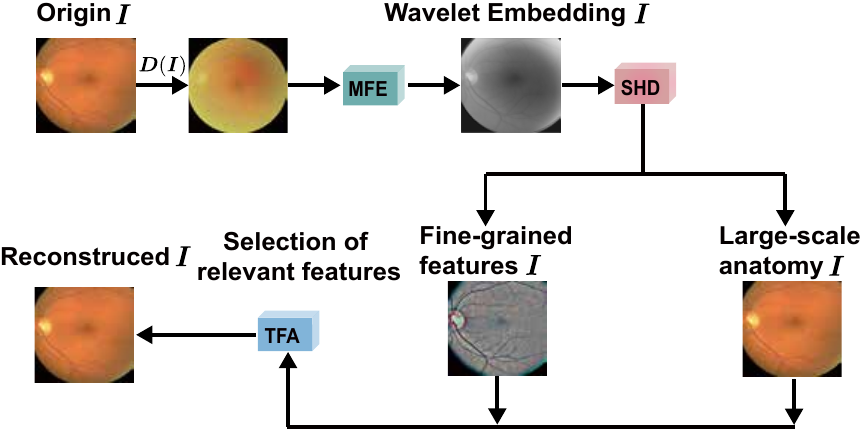}}
\caption{Overview of the proposed \ourmodel{}. The \encoder{} maps the degraded fundus image into a wavelet feature pyramid. The \decoder{} reconstructs two complementary streams: fine-grained, high-frequency details and large-scale anatomical structures. The \attention{} applies spatial–channel attention to select and fuse the most relevant signals from both streams to produce the final reconstruction.
}
\label{fig:brief}
\vspace{-0.5em}
\end{figure}

Our learning framework is illustrated in Fig. \ref{fig:brief}. The \encoder{} progressively extracts features at different spatial resolutions, while two complementary decoding paths are designed to preserve information at different levels. The high-frequency path focuses on reconstructing fine-grained structures such as vessels and small lesions, whereas the origin-domain decoding path ensures the consistency of large-scale anatomical structures such as the optic disc and macula. On this basis, the \attention{} module employs joint spatial and channel attention to adaptively select the most relevant features from the two paths.

\begin{figure*}[t]
\centerline{\includegraphics[width=0.95\textwidth]{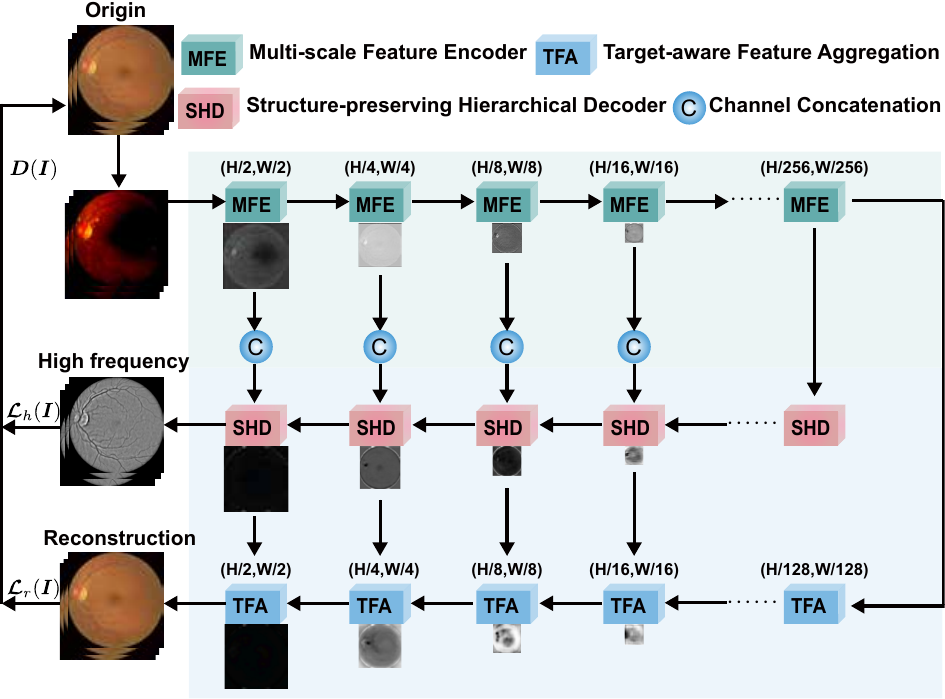}
}
\caption{The workflow of the \ourmodel{} framework. First, a high-quality image $I$ is degraded using a transformation function $D(I)$ to generate a low-quality input. The degraded image is processed by a multi-level encoder that progressively downsamples the input, generating an embedded representation enriched with multi-level wavelet features. This representation is then upsampled by a series of high-frequency convolutional decoders, and supervision loss is computed by comparing the recovered and original high-frequency components. At each resolution scale, an attention-based fusion module combines the wavelet-encoded features with the upsampled high-frequency representations, further enhancing fine details. Finally, the high-frequency-enhanced embedding, which integrates multi-level wavelet features, is used to reconstruct the high-quality image.}
\label{fig:overview}

\end{figure*}

Specifically, a high-quality image $I$ is first degraded by a transformation $D(I)$, then processed by the \encoder{} module.
The \encoder{} extracts four sub-band features: approximate, vertical, horizontal, and diagonal components. 
These sub-bands are refined using depthwise separable convolutions, which enhance feature extraction efficiency while preserving critical multi-scale structural details. 
The resulting multi-scale features are then progressively upsampled by the \decoder{}. 
Each \decoder{} block incorporates group attention and depthwise separable convolutions to integrate multi-scale information and accurately restore high-frequency details. 
To further enhance high-frequency information, the upsampled features at each scale are fused with encoder embeddings via the \attention{} module. 
The final embedding, enriched with high-frequency details, is employed to reconstruct the enhanced high-quality image. The detail is illustrated in 
Fig. \ref{fig:overview}.

\subsection{Multi-scale Feature Encoder}

\begin{figure}[t]
\centerline{\includegraphics[width=0.98\columnwidth]{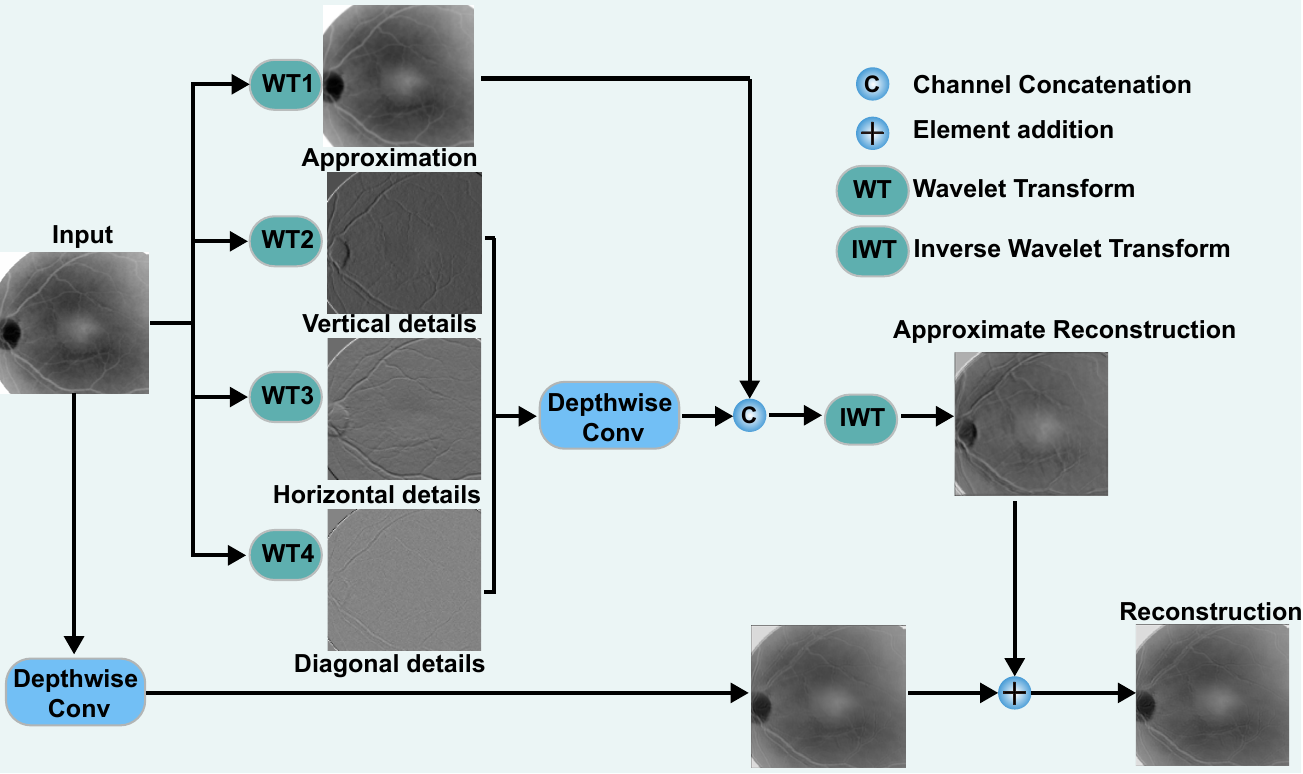}}
\caption{The high-frequency enhanced encoder decomposes the input image into multiple hierarchical frequency sub-bands. Using downsampling and wavelet transforms, the \encoder{} extracts four components: approximation, vertical, horizontal, and diagonal. These sub-band features are refined through depthwise separable convolutions and then reconstructed via inverse wavelet transform, enabling robust embedding of both coarse and fine structural details. A residual connection, implemented with depthwise separable convolutions, links the output back to the input. This prepares the multi-scale features for attention-based fusion and high-frequency enhancement in the decoding stage.
}
\label{fig:MFE}
\vspace{-0.5em}
\end{figure}

The \encoder{} is designed to extract robust multi-scale features by employing a hierarchical decomposition and reconstruction strategy, enabling more effective downstream image reconstruction. By applying progressive frequency decomposition and embedding techniques, it captures both global context and fine-grained details of the fundus images.
As illustrated in Fig. \ref{fig:MFE}, the \encoder{} applies a WT to the input features, decomposing them into four sub-band components: one global component and three detail components that represent vertical, horizontal, and diagonal structures, respectively. 
The WT decomposition is implemented with predefined convolution kernels, formulated as:
\begin{align}
f_g = &\frac{1}{2}{
\begin{bmatrix}
1 & \quad \ \, 1 \\
1 & \quad \ \, 1
\end{bmatrix}}, \;
f_{d_1}= \frac{1}{2}{
\begin{bmatrix}
1 & \ \ -1 \\
1 & \ \ -1
\end{bmatrix}},
\notag
\\
f_{d_2}= &\frac{1}{2}{
\begin{bmatrix}
1 &  1 \\
-1 & -1
\end{bmatrix}}, \;
f_{d_3}=\frac{1}{2}{
\begin{bmatrix}
1 &  -1 \\
-1 &  -1
\end{bmatrix}},
\end{align}

Given an input feature map $X$, we can obtain a set of sub-band components comprising one global and three directional detail components:
\begin{equation}\{X_g,X_{d_1},X_{d_2},X_{d_3}\}=f_{WT}([f_g,f_{d_1},f_{d_2},f_{d_3}],X).\end{equation}
Here, $X_g$ captures the global (low-frequency) information.
$X_{d_1}$, $X_{d_2}$, and $X_{d_3}$ capture detail features in the vertical, horizontal, and diagonal directions, respectively. 
The detail components, denoted as $X_D$, are processed via depthwise separable convolutions to enhance feature selectivity and representation capability while maintaining computational efficiency:
\begin{equation}X_{D_o}=f_{DWC}(X_D),\end{equation}
where $f_{DWC}$ denotes the depthwise separable convolution operation. The refined detail features $X_{D_o}$ are concatenated with the global component $X_g$ along the channel dimension to form the integrated embedded feature $X_e$:
\begin{equation}
X_e = [X_g, X_{D_o}].
\end{equation}
The integrated feature $X_e$ is then transformed back into the spatial domain via the inverse wavelet transform (IWT), yielding $X_r$, which fuses global and detail information:
\begin{equation}X_r=f_{IWT}(X_e).\end{equation}
To further enhance representation, $X_r$ is element-wise added to the output of a depthwise separable convolution applied directly to $X$:
\begin{equation}X_o=X_r+f_{DWC}(X).\end{equation}
The operation of the $l$-th layer in the \encoder{} can thus be defined as: 
\begin{equation}X_{enc}^l=IWT\left([X_g^l,f_{DWC}(X_D^l)]\right)+f_{DWC}(X_{enc}^{l-1}),\end{equation}
where $X_{enc}^l$ denotes the output of the $l$-th encoder layer, with $l = 1, 2, \ldots, L$, and $L$ representing the total number of layers.

\subsection{Structure-preserving Hierarchical Decoder}

\begin{figure}[t]
\centerline{\includegraphics[width=\columnwidth]{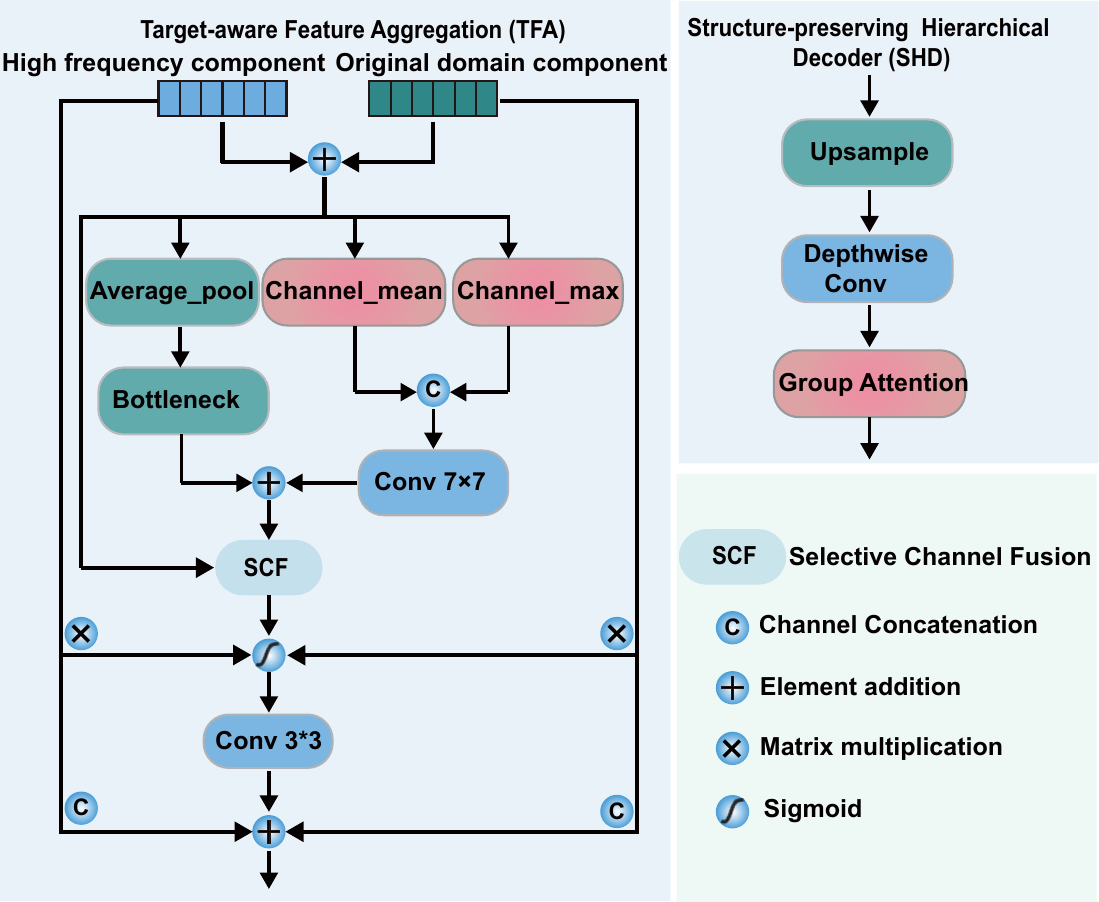}}
\caption{The unified attention fusion module combines spatial and channel information using spatial average pooling, along with channel-wise mean and max operations. A bottleneck structure followed by convolutional layers further enhances feature selectivity. The SCF module refines the fused features, emphasizing important high-frequency components for improved decoding performance.
The high-frequency convolutional decoder progressively upsamples the multi-scale wavelet features. Leveraging group attention and depthwise separable convolutions, the \decoder{} reconstructs high-quality images with minimal parameter overhead, while effectively preserving fine-grained, high-frequency details.
}
\label{fig:TFA and SHD}
\vspace{-0.5em}
\end{figure}

The decoder is designed to progressively upsample the encoder-generated features while adaptively fusing high-frequency details with global structural information. As shown in Fig. \ref{fig:TFA and SHD}, it comprises two key modules: \decoder{} and \attention{}.

The \decoder{} module restores low-resolution features to high-resolution outputs through efficient upsampling, while preserving essential structural characteristics.
Depthwise separable convolutions combined with channel shuffle are employed to enable efficient and lightweight upsampling. Given the $({l-1})^{\text{th}}$ level decoder input $X_{dec}$, the upsampling operation is defined as:
\begin{equation}
X_{up}^l=f_{DWC}(f_{Upsample}(X_{dec}^{l-1})).
\end{equation}
To further improve feature expressiveness, a group attention mechanism is applied to the upsampled feature map:
\begin{equation}X_{ga}^l=f_{GA}(X_{up}^l,G),\end{equation}
where $G = {C_1, C_2, \ldots, C_n}$ defines the channel groups, and $f_{GA}$ represents the Group Attention function, which enhances inter-group interactions using channel shuffle and pointwise convolutions.

\ourmodel{} employs two complementary supervision paths to jointly constrain the enhancement: the first is high-frequency supervision, which aims to preserve fine-grained structures, and the second is reconstruction supervision, which enforces global appearance and large-scale anatomical consistency. 
At each scale, the \attention{} module applies joint spatial–channel attention. 
Spatial attention, guided by high-frequency supervision maps, highlights vessels and small lesions in fine-grained regions. 
In parallel, channel attention adaptively amplifies frequency bands or channels relevant to diagnosis (e.g., texture and edge channels) while suppressing irrelevant responses.
Meanwhile, the reconstruction branch enforces consistency with the input in large-scale structures (e.g., optic disc and macular regions), preventing artifact introduction from over-sharpening. Through the selection mechanism of \attention{}, the two supervision paths achieve a “detail–structure” complementarity.

The \attention{} module first computes spatial and channel attention maps:
\begin{equation}
I_{sa} = \sigma\left(\mathrm{Conv}_{7\times7}\left([f_{AP}(I_{in}), f_{MP}(I_{in})]\right)\right),
\end{equation}
where $\sigma$ denotes the sigmoid function, and $\mathrm{Conv}_{7\times7}$ is a 7×7 convolution. Here, $f_{AP}$ and $f_{MP}$ denote average pooling and max pooling, respectively.
\begin{equation}
I_{ca}=\sigma\left(f_{PWC}\left(\mathrm{ReLU}\left(f_{PWC}\left(f_{GAP}(I_{in})\right)\right)\right)\right),
\end{equation}
where $f_{GAP}$ denotes global average pooling, and $f_{PWC}$ is the pointwise convolution operation.

Next, spatial and channel attention are fused in the Selective Channel Fusion (SCF) module:
\begin{equation}I_{pa}=\sigma\left(\mathrm{Conv}_{re}\left([I_{sa},I_{ca}]\right)\right),\end{equation}
where $\mathrm{Conv}_{re}$ is a convolution layer with reflective padding.
Finally, a step-by-step fusion is performed at each decoding scale:
\begin{equation}I_{A_o}^l=\mathrm{Conv}_{1\times1}\left(I_{pa}\odot I_{h_o}^l\right)+(1-I_{pa})\odot X_{ga},\end{equation}
where $\odot$ denotes element-wise multiplication.

\subsection{Loss Function}

The loss function of \ourmodel{} is designed to jointly supervise high-frequency enhancement and global reconstruction quality. The decoder consists of two branches: one dedicated to enhancing high-frequency details, and the other to reconstructing the global image. The final output quality is improved through the joint optimization of both branches. The loss components are defined as follows:

The high-frequency enhancement loss measures the difference between the high-frequency components of the generated image and the ground truth, extracted via a Gaussian high-pass filter. This loss optimizes decoder path 1 for capturing fine details.
\begin{equation}\mathcal{L}_h=\left\|P_h-G_h(P_g)\right\|_1.\end{equation}
Here, $P_h$ is the high-frequency enhanced output from decoder path 1, $G_h$ is the Gaussian high-pass filter operator, and $P_g$ is the ground truth high-quality image.

\begin{equation}\mathcal{L}_r\mathrm= \|P_r-P_g\|_2,\end{equation}
where $P_r$ is the reconstructed image produced by decoder path 2.

The total loss $\mathcal{L}_t$ combines both components using a weighted sum, balancing high-frequency fidelity with global reconstruction accuracy:
\begin{equation}
\mathcal{L}_t={ \lambda}\mathcal{L}_h+(1-\lambda)\mathcal{L}_r,
\label{eq:16}
\end{equation}
where $\lambda \in [0, 1]$ controls the trade-off between high-frequency detail preservation and overall image fidelity.

\section{Experiments and results}

\subsection{Datasets}

We evaluated the performance of our algorithm using five benchmark datasets. The five datasets used are listed below:
\begin{itemize}
\item  \textbf{BA}
{was collected by PCSIR, Karachi under their project of ``Development of digital fundus images dataset for Artificial Intelligence based model development and performance testing''.
The retinal fundus images were acquired in collaboration with Al-Ibrahim Hospital, Karachi, Pakistan.
It contains 100 high quality image pairs acquired from 5 doctor’s repeated fundus examination.}

\item  \textbf{Drive  \citep{staal2004ridge}}
{was introduced as part of a retinal screening program in the Netherlands to support research in automated retinal vessel segmentation. 
It contains 40 high-resolution color fundus images, each accompanied by expert-annotated vessel segmentation maps, providing a reliable ground truth for model evaluation.
}

\item  \textbf{Kaggle  \citep{kaggle2024diabetic}}
{comprises high-resolution retinal images captured under diverse imaging conditions from multiple clinical sources. Each image is labeled according to the International Clinical Diabetic Retinopathy Severity Scale, with grades ranging from 0 (No DR) to 4 (Proliferative DR).}

\item  \textbf{EyeQ  \citep{nguyen2020diabetic}}
{consists of 28,792 retinal images categorized into three quality levels: `Good', `Usable', and `Reject'. For our experiments, we selected 5,000 high-quality images labeled as `Good'.}

\item  \textbf{Refuge  \citep{orlando2020refuge}}
{is a benchmark dataset designed for evaluating automated glaucoma detection methods. It includes fundus images with detailed annotations of the optic disc and optic cup. A total of 400 images were selected for use in our study.}
\end{itemize}

\subsection{Data Pre-processing}

We developed a multi-stage image degradation pipeline to simulate common quality impairments in fundus images, including variations in brightness, contrast, and saturation, as well as halo artifacts, dark spots, noise, and blur. The method is built upon three primary degradation transformations that collectively model the diverse artifacts observed in real-world fundus imaging.

Brightness and contrast variations are simulated by adjusting image brightness, contrast, and saturation to reflect changes caused by different imaging devices or lighting environments. Randomly distributed circular noise is added to emulate sensor-induced artifacts and noise commonly introduced by imaging hardware. Gaussian blur is used to mimic focus deviations and optical imperfections, with varying kernel sizes and orientations to represent different levels of blurriness. Cataract degradation simulates the visual distortions experienced by cataract patients. It begins with globally applied Gaussian blur using randomly sampled parameters. A radial distance map centered in the image is then generated to create a halo effect, with intensity modulated to accentuate cataract-like distortions. Finally, the image undergoes fine-tuning of contrast, brightness, and color balance to enhance realism.

\begin{figure}[t]
\centerline{\includegraphics[width=\columnwidth]{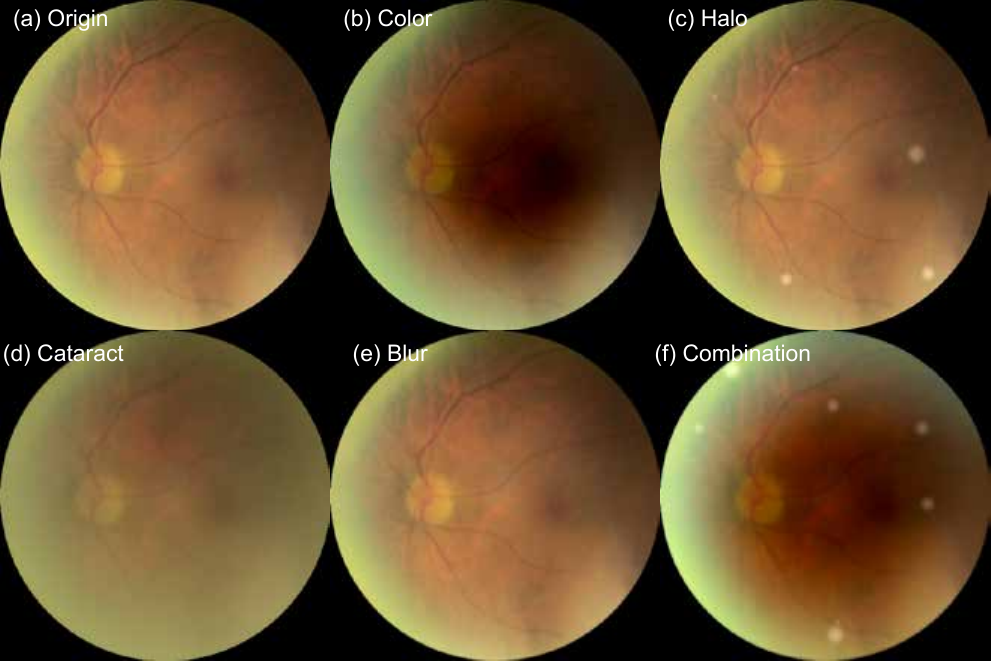}}
\caption{Images synthesized on the BA dataset using different degradation methods. (a) Original high-quality image; (b) Image degraded with brightness and saturation adjustments; (c) Image with added circular spot noise; (d) Image affected by simulated cataract degradation; (e) Image with applied Gaussian blur; (f) Image with multiple combined degradation types.
}
\label{fig:noise}
\vspace{-2em}
\end{figure}

These basic transformations are permuted and combined to simulate more complex and realistic degradation scenarios. Representative examples of images with different simulated degradations are shown in Fig. \ref{fig:noise}.

For fair comparison with methods using synthetic degradations, each test image was subjected to eight types of degradation during evaluation. These degradations were randomly generated combinations of four base types: illumination variation, noise, blur, and cataract simulation. Each dataset was divided into training and testing sets in a 7:3 ratio, ensuring all available images were used. 

\subsection{Comparison methods}

\begin{itemize}

    \item \textbf{DCP}  \citep{he2010single} is a traditional denoising method based on reflection and illumination decomposition. Its performance depends on global image statistics and handcrafted mapping functions.

    \item \textbf{Cofe-Net}  \citep{shen2020modeling} utilizes paired synthetic images and incorporates vessel segmentation to preserve retinal structures during image enhancement.

    \item \textbf{StillGAN}  \citep{ma2021structure} is a GAN-based method using unpaired supervision, similar to CycleGAN  \citep{zhu2017unpaired}. It demonstrates strong signal intensity in the reconstructed images.
            
    \item \textbf{ArcNet}  \citep{li2022annotation} is a CNN-based approach that accesses test data during training to adaptively generate models suited for real-world deployment.

    \item \textbf{RFormer}  \citep{deng2022rformer} is a GAN-based method trained on clinically collected high- and low-quality image pairs. It employs a Swin Transformer-like backbone for enhanced feature extraction.
    
    \item \textbf{GfeNet}  \citep{li2023generic} is a self-supervised model with a large parameter size. It integrates high-frequency supervision to enhance detail preservation.

\end{itemize}

\subsection{Experimental settings}
All experiments were implemented in PyTorch 2.1 and conducted on a single NVIDIA A100 GPU with 40GB of memory. We used the AdamW optimizer with a linear learning rate decay schedule. The initial learning rate was set to 0.0002, with decay starting in the final 50 epochs. The optimizer's momentum parameter was set to 0.5. All labeled and unlabeled images are resized to $512 \times 512$. The random seed is fixed at 42, and the batch size is set to 16.

To comprehensively evaluate image enhancement and downstream task performance, we adopt several widely used quantitative metrics. Specifically, Structural Similarity Index (SSIM) and Peak Signal-to-Noise Ratio (PSNR) are employed to measure image quality, reflecting structural fidelity and signal reconstruction accuracy, respectively. For vessel segmentation, we use Intersection over Union (IoU) and Dice coefficient, which quantify the overlap between predicted and ground-truth vessel masks, as well as Area Under Curve (AUC) to evaluate overall segmentation robustness. For optic disc/cup detection, we report mean Intersection over Union (mIoU) and Dice coefficient to assess the accuracy of anatomical structure localization.

To provide more rigorous comparisons, we report each metric as mean $\pm$ standard deviation on all test images. Furthermore, we performed paired statistical significance tests between our method and the strongest baseline (chosen based on SSIM performance). Specifically, paired t-tests were performed and Wilcoxon signed-rank tests were additionally used to validate robustness. Statistical significance levels are denoted by superscripts in Table~\ref{tab:results}, where $^{*}$, $^{**}$, and $^{***}$ indicate $p<0.05$, $p<0.01$, and $p<0.001$, respectively.
\begin{table*}[t]
    \centering
    \caption{Comparison of image enhancement quality across different methods on five datasets of varying scales.}
    \setlength{\tabcolsep}{1.2mm}{
    \begin{tabular}{lccccc}
    \toprule
       \multirow{2}{*}{\textbf{Methods}} & BA & DRIVE & Kaggle & Subset-EyeQ & Refuge \\ 
        ~ & SSIM / PSNR & SSIM / PSNR & SSIM / PSNR & SSIM / PSNR & SSIM / PSNR \\ 
    \midrule
        DCP \citep{he2010single} & \makecell{0.864 ± 0.082 \\ 20.58 ± 8.08} & \makecell{0.885 ± 0.075 \\ 19.36 ± 5.80} & \makecell{0.867 ± 0.072 \\ 18.05 ± 4.15} & \makecell{0.846 ± 0.082\\ 18.31 ± 4.72} & \makecell{0.815 ± 0.075 \\ 16.23 ± 2.94} \\ 
        Cofe-Net \citep{shen2020modeling} & \makecell{0.910 ± 0.057\\ 20.54 ± 3.31} & \makecell{0.930 ± 0.029 \\ 22.10 ± 4.18} & \makecell{0.949 ± 0.030 \\ 25.61 ± 4.34} & \makecell{0.919 ± 0.042 \\ 24.87 ± 4.51} & \makecell{0.933 ± 0.031 \\ 25.82 ± 3.55} \\ 
        StillGAN \citep{ma2021structure} & \makecell{0.913 ± 0.054 \\ 23.45 ± 5.90} & \makecell{0.944 ± 0.040 \\ 27.91 ± 6.53} & \makecell{0.947 ± 0.035 \\ 28.56 ± 9.31} & \makecell{0.931 ± 0.049 \\ 28.08 ± 8.63} & \makecell{0.925 ± 0.051 \\ 29.75 ± 8.52} \\
        ArcNet \citep{li2022annotation} & \makecell{0.903 ± 0.042 \\ 19.87 ± 3.55} & \makecell{0.943 ± 0.032 \\ 23.70 ± 2.56} & \makecell{0.921 ± 0.028 \\ 21.99  ± 2.76} & \makecell{0.900 ± 0.040 \\ 22.17 ± 2.98} & \makecell{0.893 ± 0.040 \\ 21.79 ± 2.48} \\ 
        RFormer \citep{deng2022rformer} & \makecell{0.906 ± 0.056 \\ 23.64 ± 7.40} & \makecell{0.925 ± 0.039 \\ 26.20 ± 5.04} & \makecell{0.926 ± 0.0483 \\ 25.73 ± 6.70} & \makecell{0.899 ± 0.064 \\ 23.57 ± 6.68} & \makecell{0.912 ± 0.046 \\ 25.08 ± 5.28} \\
        GfeNet \citep{li2023generic} & \makecell{0.928 ± 0.029 \\ 22.65 ± 2.53} & \makecell{0.946 ± 0.026 \\ 25.59 ± 3.61} & \makecell{0.957 ± 0.020 \\ 27.10 ± 3.53} & \makecell{0.944 ± 0.034 \\ 27.10 ± 2.90} & \makecell{0.935 ± 0.040 \\ 27.22 ± 3.11} \\ 
        \ourmodel{} & \makecell{\textbf{0.933 ± 0.024\textsuperscript{**}} \\ 22.97 ± 3.42} & \makecell{\textbf{0.950 ± 0.025\textsuperscript{***}} \\ 26.05 ± 4.36} & \makecell{\textbf{0.963 ± 0.019\textsuperscript{***}} \\ \textbf{30.49 ± 5.52}\textsuperscript{***}} & \makecell{\textbf{0.947 ± 0.029} \\ \textbf{28.13 ± 4.31}\textsuperscript{***}} & \makecell{\textbf{0.950 ± 0.025\textsuperscript{***}} \\ 29.05 ± 3.41\textsuperscript{***}} \\ 
    \bottomrule
    \end{tabular}  
    }
    \label{tab:results}    
\end{table*}

\begin{table*}[!ht]
    \centering
    \caption{Comparison of representative image enhancement methods. The table summarizes their training modes, types of training data, and the number of parameters.}
    \label{tab:comparison}   
    \begin{tabular}{lllc}
    \toprule
        Methods & Training mode & Training data & Number of Parameters (M) \\
    \midrule
        DCP  \citep{he2010single} & No training & ~ & ~ \\
        Cofe-Net  \citep{shen2020modeling} & Additional  supervisory tags & High-quality images & 54.42 \\
        StillGAN  \citep{ma2021structure} & Additional  supervisory tags & High-quality images & 21.11 \\ 
        ArcNet  \citep{li2022annotation} & Additional supervisory tags & High-quality images & 78.64 \\ 
        RFormer  \citep{deng2022rformer} & Supervised training & High and low quality images & 41.22 \\ 
        GfeNet  \citep{li2023generic} & self-supervised training & High-quality images & 89.3 \\ 
        \ourmodel{} & self-supervised training & High-quality images & 6.95 \\ 
     \bottomrule   
    \end{tabular}
    
\end{table*}

\begin{figure*}[t]
    \centering
    \includegraphics[width=0.98\textwidth]{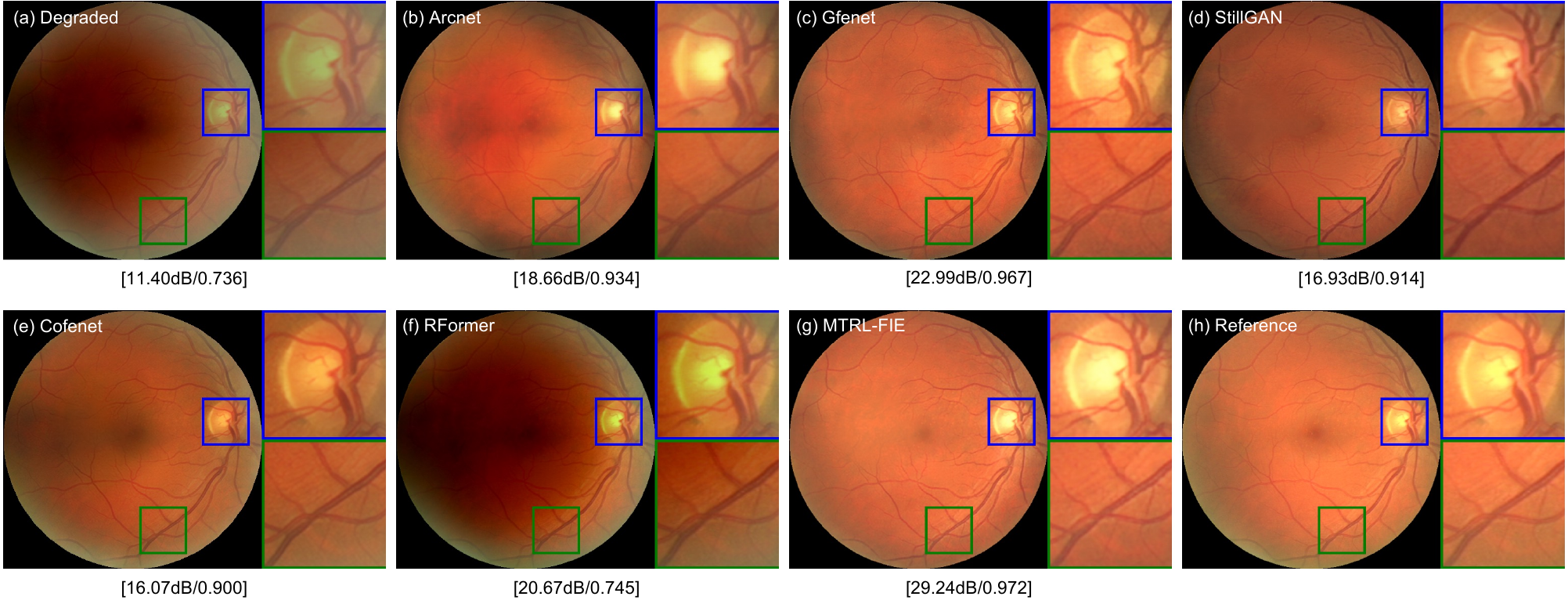}    
    \caption{Visual comparison of enhancement results. (a) Synthetic low-quality fundus image. (b–g) Enhancement results by different methods. (h) The ground truth high-quality image. The notation [. / .] indicates the corresponding [PSNR / SSIM] scores. The right section of each image highlights visual differences in the optic disc/cup and blood vessel regions.    
    }
    \label{fig:enhancement}
    \vspace{-0.5em}
\end{figure*}

\subsection{Performance comparison}

\subsubsection{Fundus image enhancement}
Table \ref{tab:results} quantitatively compares the enhancement performance of different methods across five datasets using SSIM and PSNR. Traditional method DCP shows limited improvements due to its reliance on handcrafted priors, while supervised methods such as Cofe-Net and RFormer achieve relatively higher scores but struggle with generalization across datasets. GAN-based approaches like StillGAN provide better structural recovery but often introduce artifacts or insufficient contrast restoration. GfeNet demonstrates strong performance, particularly in SSIM. In contrast, our proposed MTRL-FIE consistently achieves the highest SSIM and PSNR on most datasets, with statistically significant improvements (p < 0.01 or p < 0.001 in several cases). 

To facilitate visual comparison of enhancement performance,we present the synthesized low-quality fundus images alongside the results produced by various baseline methods. As shown in Fig. \ref{fig:enhancement}, RFormer  \citep{deng2022rformer} and Cofe-Net  \citep{shen2020modeling} fail to effectively correct lighting degradation in the optic disc and optic cup regions. StillGAN  \citep{ma2021structure} struggles with generating vascular structures, and irregular artifacts are visible in the optic cup area. While ArcNet yields relatively accurate structural representations of the optic disc and cup, it performs poorly in restoring global image saturation. GfeNet  \citep{li2023generic} exhibits slight distortion in the vessel details due to residual lighting artifacts. In contrast, \ourmodel{} produces enhanced images with more natural brightness and better-preserved anatomical structures, particularly in critical regions such as the optic disc, optic cup, and retinal vasculature.

We also compared the number of training parameters between our proposed \ourmodel{} and several baseline methods, as summarized in Table~\ref{tab:comparison}. As shown, GfeNet \citep{li2023generic}, ArcNet \citep{li2022annotation}, Cofe-Net \citep{shen2020modeling}, RFormer \citep{deng2022rformer}, and StillGAN \citep{ma2021structure} all require significantly more parameters than our model. 
\subsubsection{Vessel segmentation}
\begin{figure*}[t]
\centering
\includegraphics[width=0.98\textwidth]{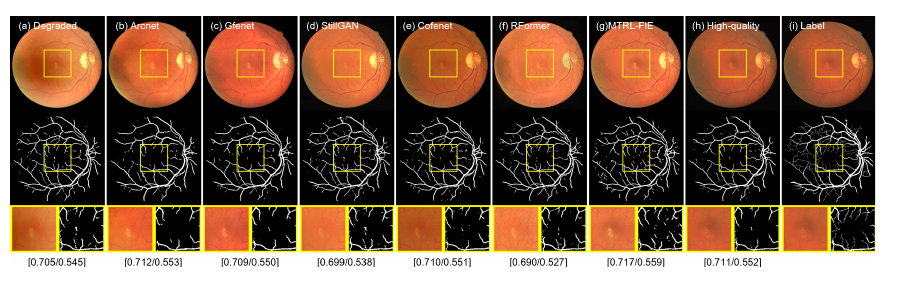}
\caption{Visual comparison of vessel segmentation results after enhancement on synthetic low-quality fundus images. From top to bottom, each column displays the enhanced image, the corresponding vessel segmentation result, and a zoomed-in view of a key region. (a) The synthetic low-quality image; (b–g) The results from baseline enhancement methods; (h) The segmentation result using the original high-quality image; (i) The ground truth vessel annotation from the DRIVE  \citep{staal2004ridge} dataset.}
\label{fig:segementation}
\end{figure*}

To assess the effectiveness of image enhancement, we evaluated various methods on a downstream vessel segmentation task using the degraded DRIVE \citep{staal2004ridge} test set. Quantitative results (Table \ref{tab:segementation}) demonstrate that our framework achieves superior performance across AUC, IoU, and Dice coefficient metrics.
We employed a U-Net \citep{cardoso2022monai} model trained on the DRIVE \citep{staal2004ridge} training set for segmentation. The segmentation outputs on synthesized low-quality fundus images generated by various enhancement methods are illustrated in Fig \ref{fig:segementation}. Our method more effectively preserves vascular structures.
The baseline method DCP \citep{he2010single}, which primarily enhances global contrast and lighting, struggles to correct localized artifacts like light spots, leading to inferior segmentation performance. StillGAN \citep{ma2021structure} and similar approaches use various loss functions to reconstruct overall image structure. However, they overlook retinal features critical for diagnosis, resulting in the loss of fine vascular details, especially endpoints. Cofe-Net \citep{shen2020modeling} incorporates pre-trained segmentation models to enhance retinal feature representation and anatomical accuracy. 
Whereas ArcNet \citep{li2022annotation} and GfeNet \citep{li2023generic} prioritize global image quality in their loss functions and lack dedicated retinal feature encoders, limiting their ability to recover detailed vascular structures. 
RFormer \citep{deng2022rformer}, which employs a Transformer-based encoder, faces difficulties in generating continuous vascular networks due to the absence of the local inductive bias offered by convolutional layers.

\begin{table}[!ht]
    \centering    
    \caption{Comparison of vessel segmentation performance across different methods on the DRIVE dataset.}
    \label{tab:segementation}
    \setlength{\tabcolsep}{2.5mm}{
    \begin{tabular}{lccc}
    \toprule
        Methods & IoU & Dice & AUC \\ 
    \midrule
        Low-quality  & 0.527 & 0.684 & 0.848 \\ 
        High-quality  & 0.575 & 0.73 & 0.884 \\ 
        GfeNet  \citep{li2023generic} & 0.541 & 0.7 & \textbf{0.865} \\ 
        ArcNet  \citep{li2022annotation} & 0.532 & 0.689 & 0.85 \\ 
        Cofe-Net  \citep{shen2020modeling} & 0.543 & 0.7 & 0.86 \\ 
        RFormer  \citep{deng2022rformer} & 0.495 & 0.657 & 0.84 \\ 
        DCP  \citep{he2010single} & 0.493 & 0.655 & 0.834 \\ 
        StillGAN  \citep{ma2021structure} & 0.525 & 0.684 & 0.843 \\ 
        \ourmodel{} & \textbf{0.546} & \textbf{0.704} & \textbf{0.865} \\ 
    \bottomrule
    \end{tabular}
    }  
    \vspace{-1.2em}
\end{table}

\begin{table}[!ht]
    \centering
    \caption{Comparison of different enhancement methods for optic disk and cup detection performance on the Refuge  \citep{orlando2020refuge} dataset.}
    \label{tab:detection}
    \setlength{\tabcolsep}{4mm}{
    \begin{tabular}{lccc}
    \toprule
        Methods & mIoU & Dice \\ 
    \midrule
        Low-quality  & 0.737 & 0.842 \\ 
        High-quality  & 0.788 & 0.88 \\ 
        GfeNet  \citep{li2023generic} & 0.769 & 0.868 \\ 
        ArcNet  \citep{li2022annotation} & 0.782 & 0.876 \\ 
        Cofe-Net  \citep{shen2020modeling} & 0.769 & 0.868 \\ 
        RFormer  \citep{deng2022rformer} & 0.718 & 0.824 \\ 
        DCP  \citep{he2010single} & 0.708 & 0.822 \\ 
        StillGAN  \citep{ma2021structure} & 0.785 & 0.878 \\ 
        \ourmodel{} & \textbf{0.789} & \textbf{0.881} \\ 
    \bottomrule
    \end{tabular}
     \vspace{-1.3em}
    }   
\end{table}
In contrast, our \encoder{} integrates multi-scale learning techniques to extract more effective and robust structural features. Furthermore, the decoder selectively retains the most relevant information, enhancing the quality of the reconstructed images. The images generated by \ourmodel{} outperform the original inputs, especially in the context of vessel segmentation. Consequently, our high-quality enhanced images achieve the best segmentation performance, exceeding all baseline methods.

\subsubsection{Optic disc/optic cup detection}
\begin{figure*}[t]
\centering
\includegraphics[width=0.98\textwidth]{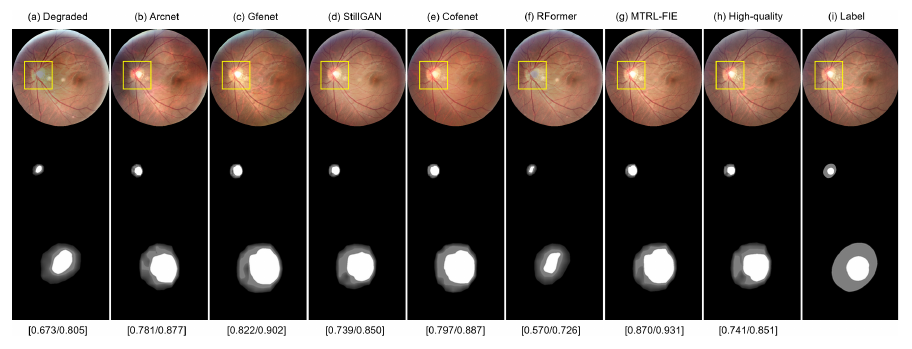}
\caption{Optic disc and cup detection results on enhanced synthetic low-quality fundus images. From top to bottom: the enhanced image, the corresponding detection maps, and magnified views. (a) The synthetic low-quality image; (b–g) Results from comparison methods; (h) Displays the segmentation using the original high-quality image; (i) The ground truth annotation from the Refuge \citep{orlando2020refuge} dataset.}
\label{fig:detection}
\end{figure*}

To evaluate enhancement effectiveness on optic disc and cup detection, we tested various methods using the degraded Refuge  \citep{kaggle2024diabetic} test set. Specifically, we degraded the fundus images, applied different enhancement models, and used a unified PraNet  \citep{fan2020pranet} model for detection. The results are presented in Fig. \ref{fig:detection} and Table \ref{tab:detection}, where \ourmodel{} consistently achieved superior performance in terms of mIoU and Dice coefficient.

Despite the relatively high quality of the original images, traditional enhancement methods struggle to handle multi-scale and multi-frequency information, leading in poor detail preservation and structural fidelity. In contrast, \ourmodel{} effectively enhances fine details while preserving global structures, yielding stronger performance in the optic disc and cup detection task. Remarkably, it even surpasses the original unenhanced images. These results highlight the effectiveness of our framework in both image enhancement and downstream analysis, especially in retaining critical visual details and anatomical structures.

\begin{table}
	\centering	
        \caption{Ablation study of the proposed modules in BA dataset.}
        \setlength{\tabcolsep}{3mm}{
	\begin{tabular}{lccccc} 
		\toprule 
		\multirow{2}{*}{Configuration} & \multicolumn{2}{c}{Modules}&\multicolumn{2}{c}{Metrics}
		\\ & ~\encoder{}&\decoder{}& SSIM&PSNR \\
		\midrule
		w/o \encoder{}-\decoder{}& \XSolidBrush & \XSolidBrush & 0.868&20.51 \\
		w/o \decoder{}& \XSolidBrush & \Checkmark & 0.915&22.62\\
		w/o \encoder{}& \Checkmark & \XSolidBrush & 0.925&22.64 \\
		Our framework & \Checkmark &  \Checkmark & \textbf{0.933} & \textbf{22.97} \\
		\bottomrule
	\end{tabular}
	\label{tab:ablation}       
        }
 \vspace{-1em}
\end{table}

\begin{table}[t]
    \centering
    \caption{Ablation study on the effect of $\lambda$ in BA dataset.}
    \setlength{\tabcolsep}{6mm}{
    \begin{tabular}{lcc}
        \toprule
        $\lambda$ & SSIM & PSNR \\
        \midrule
        0.75 & 0.927 ± 0.023 & 22.65 ± 3.42 \\
        0.67 & \textbf{0.933 ± 0.024} & \textbf{22.97 ± 3.42} \\
        0.50 & 0.933 ± 0.025 & 22.86 ± 3.50 \\
        0.33 & 0.914 ± 0.030 & 21.73 ± 3.16 \\
        0.25 & 0.918 ± 0.026 & 21.96 ± 3.36 \\
        \bottomrule
    \end{tabular}}
    \label{tab:lambda}
    \vspace{-1.5em}
\end{table}

\subsection{Ablation Study}

To assess the contribution of each component in \ourmodel{}, we performed an ablation experiment on the BA dataset. The results are summarized in Table \ref{tab:ablation}. We started with a baseline model using only a standard U-Net. This configuration yielded low SSIM and PSNR scores, reflecting its limited ability to capture global context and restore high-frequency details. Adding the \encoder{}, which utilizes wavelet transforms to extract hierarchical features, notably improved performance—especially in PSNR—highlighting its effectiveness in capturing global structures and fine details. When incorporating only \decoder{}, we observed a significant increase in SSIM. This module improves restoration quality by providing target-aware features guidance, particularly for retinal structures. Finally, combining both \encoder{} and \decoder{} in the full \ourmodel{} architecture led to the highest performance across both PSNR and SSIM, confirming the complementary benefits of global and fine-detail restoration.

In addition, we also investigate the effect of the trade-off parameter $\lambda$, which governs the balance between the high-frequency enhancement loss and the origin-domain reconstruction loss in Eq.\ref{eq:16}, we conducted a systematic ablation study on the BA dataset. We evaluated values of $\lambda$ corresponding to key loss ratios of 3:1, 2:1, 1:1, 1:2, and 1:3 (i.e., $\lambda = {0.75, 0.67, 0.50, 0.33, 0.25}$), as shown in Table~\ref{tab:lambda}.

The results demonstrate that both SSIM and PSNR are sensitive to the choice of $\lambda$. In particular, when $\lambda=0.67$ (2:1 ratio), the model achieves the best balance, yielding the highest SSIM ($0.933 \pm 0.024$) and PSNR ($22.97 \pm 3.42$). Therefore, we adopt $\lambda=0.67$ as the default setting in all experiments.

\section{Discussion and Conclusion}

This paper introduces \ourmodel{}, an efficient framework for fundus image enhancement that leverages wavelet convolution to model degraded images at multiple scales. The framework enhances image detail and structural perception through multi-level upsampling and group attention mechanisms, while fine-grained attention guides the enhancement process and minimizes artifacts. \ourmodel{} effectively preserves the detail and structure of high-quality images, with particular improvements in local features and global consistency, especially in visually complex regions. In downstream tasks like vessel segmentation and optic disc/cup detection, enhanced images produced by \ourmodel{} outperform the original high-quality inputs across standard evaluation metrics. This performance gain may be attributed to the \encoder{}, which enhances multi-scale features and improves structural clarity at previously blurred boundaries, often beyond what the original high-quality image provides. Moreover, \ourmodel{} features a low parameter count, allowing it to operate efficiently while delivering high performance—making it well-suited for deployment on resource-limited embedded devices.

Despite its parameter efficiency, \ourmodel{} introduces relatively high computational complexity, particularly during training. This is mainly due to the layer-wise wavelet decomposition and reconstruction, along with the multi-stage fusion of attention mechanisms, which together increase training time. Future work may focus on streamlining the wavelet transform process and exploring lightweight alternatives to attention fusion, aiming to reduce training overhead without sacrificing performance.
For multi-modal extension, future work may explore introducing modality-specific high-frequency supervision signals to replace the unified filter. For example, structural layer boundaries in Optical Coherence Tomography (OCT) can be used to generate edge maps, while vessel probability maps in Optical Coherence Tomography Angiography (OCTA) can serve as high-frequency priors, better guiding the \attention{} module to extract target-aware features. 
To achieve real-time enhancement, we will investigate engineering optimizations such as operator fusion, mixed-precision inference, and lightweight convolutions. 
For clinical validation, we will evaluate the effectiveness of the enhanced results in two ways: (i) assessing performance gains in downstream tasks, and (ii) conducting blinded expert studies to measure diagnostic availability and detail visibility.

\section{Acknowledgements}
This work was financially supported by the National Natural Science Foundation of China (82402438, 82372147), the Science and Technology Commission of Shanghai Municipality (21ZR1465200), the Fundamental Research Funds for the Central Universities, Shanghai Municipal Education Commission, and the Research Paradigm Reform and Empowering Disciplinary Leapfrog Project. In addition, PCSIR acknowledges the support provided by INIT for AWS AI and ML services, which has enabled the development and validation of deep learning models for AI-based diagnostics.

\printcredits

\bibliographystyle{cas-model2-names}

\section*{Declaration of generative AI and AI-assisted technologies in the writing process}
During the preparation of this work the author(s) used GPT in order to improve language and readability. 
After using this service, the author(s) reviewed and edited the content as needed and take(s) 
full responsibility for the content of the publication.

\bibliography{reference}
\balance

\end{document}